\documentstyle[aps,preprint]{revtex}

\begin{document}

\title{Hadronic Effects in the Pionium Ion}

\author{ G. Rasche and A. Gashi}
\address{Institut f\"{u}r Theoretische Physik der Universit\"{a}t Z\"{u}rich,
Winterthurerstrasse 190,\\ CH-8057 Z\"{u}rich, Switzerland}
\date{submitted to Physics Letters B}
\maketitle
\begin{abstract}
The hadronic properties of the pionium ion (Coulomb bound system of three charged pions) are estimated using the results for the positronium ion $P{\! s}^{\!{\! -}}$. It turns out that the hadronic shift of the ground state energy and the lifetime of the pionium ion are approximately the same as for pionium.
\end{abstract}

  	The ($\pi^{+}\pi^{-}$) atom $A_{2\pi}$ (pionium) has become of considerable experimental and theoretical interest.  Such atoms have clearly been observed at Dubna [1] and a lower limit has been obtained for the lifetime $\tau$ of the ground state [2].  A large collaboration is engaged in an experiment at CERN to measure $\tau$ with an accuracy of about 10$\%$ [3].

	In the CERN experiment,  $A_{2\pi}$ will be produced in interactions of 24 GeV protons with target nuclei. In these reactions a certain number of Coulomb bound systems consisting of three charged pions (($\pi^{-}\pi^{+}\pi^{-}$) and ($\pi^{+}\pi^{-}\pi^{+}$)) will also be produced. We refer to these systems as the negative and positive pionium ions. Since the positive pionium ion will obviously have the same binding energy and lifetime as the negative ion we always refer to the negative ion and call it  $A^{-}_{2\pi}$ for short.
	
	Whilst the hadronic energy shift and the lifetime  $\tau$ of   $A_{2\pi}$ and their connections to the scattering lengths of the hadronic $\pi\pi$-interaction have been treated exhaustively (see e.g. [4],[5], where the references to the earlier literature can be found), the properties of  $A^{-}_{2\pi}$ have not yet been investigated. It is the aim of this note to show how the well known theoretical and experimental results for $P{\! s}^{\!{\! -}}$ (the Coulomb bound system  ($e^{-}e^{+}e^{-}$) ) can be used as a starting point to treat the binding energy and the lifetime  $\tau^{-}$ of  $A^{-}_{2\pi}$.
	
	In the quantum mechanical treatment of the Coulomb binding energy and the three particle wave function of  $A^{-}_{2\pi}$ one is faced with the same difficulties as for  $P{\! s}^{\!{\! -}}$. Analytically the problem can not be solved and one has to resort to numerical variational methods ([6],[7],[8]). 

	Since in each case only {\bf{one}} mass is involved and the {\bf{same}} charges are present, the results for  $P{\! s}^{\!{\! -}}$ can be taken over directly for  $A^{-}_{2\pi}$ by the usual dimensional arguments.
	
	From the results of [6], [7], [8], [9] for $P{\! s}^{\!{\! -}}$ we conclude that there exists only one bound state of  $A^{-}_{2\pi}$. For the $\pi^{-}$-affinity of  $A_{2\pi}$ (i.e. the binding energy of  $A^{-}_{2\pi}$ against dissociation into  $A_{2\pi}$ and a free $\pi^{-}$) we get 
\begin{equation}
  	0.3267\frac{m_{c}}{m_{e}}\; {\rm eV} =89.2\; {\rm eV},
  	\label{eq1}
\end{equation}
where $\frac{m_{c}}{m_{e}}$ is the ratio of the mass of the charged pion to the electron mass. For the binding energy of  $A^{-}_{2\pi}$ ( with respect to three free charged pions ) we therefore get
\begin{equation}
  	\frac{1}{4}m_{c} \alpha^{2} +89.2\; {\rm eV}=1858.1\; {\rm eV}+89.2\; {\rm eV}=1947.3\; {\rm eV}.
  	\label{eq2}
\end{equation}
	
	From [4] we know that in addition to this Coulombic part of the energy we have to take into account the hadronic and vacuum polarisation shifts.
	The hadronic shift $\Delta E_{HAD}$ of the ground state energy of  $A_{2\pi}$ 
 is proportional to $\rho(\pi^{+}\pi^{-})$, the probability density of $\pi^{-}$
at the position of  $\pi^{+}$. Analogously the hadronic shift  $\Delta E_{HAD}^{-}$ of the ground state energy of $A^{-}_{2\pi}$ consists of two parts. The first part is proportional to $2\rho(\pi^{-}\pi^{+}\pi^{-})$, where $\rho(\pi^{-}\pi^{+}\pi^{-})$ is the probability density of one $\pi^{-}$ at the position of $\pi^{+}$, integrated over all positions of the $\pi^{+}$ and of the other  $\pi^{-}$. The proportionality constants are the same.
It is evident from dimensional arguments that 
\begin{equation}  	\frac{2\rho(\pi^{-}\pi^{+}\pi^{-})}{\rho(\pi^{+}\pi^{-})}=\frac{2\rho(e^{-}e^{+}e^{-})}{\rho(e^{+}e^{-})}. 	
\label{eq3}
\end{equation}
The second part of the hadronic shift is proportional to the probability density of the two $\pi^{-}$ being in the same position. Because of the Coulomb repulsion this is very small compared to $\rho(\pi^{-}\pi^{+}\pi^{-})$. A simple estimate using fig. 6 of [8] shows that the ratio is $\approx 0.017$. Since the hadronic interaction in the $\pi^{-}\pi^{-}$-system is of the same order of magnitude as in the $\pi^{-}\pi^{+}$-system, this second part of $\Delta E_{HAD}^{-}$ is completely unimportant numerically.

	Anticipating the result that the right hand side of (3) is $\approx 1$, we thus have 
  \begin{equation}
\Delta E_{HAD}^{-}\approx\Delta E_{HAD},
\label{eq4}
  \end{equation}
where $\Delta E_{HAD}\approx 2.8\; {\rm eV}$ from [4], [5].

	With identical arguments one finds that 
 \begin{equation}
\tau^{-}\approx\tau,
\label{eq5}
  \end{equation}
where $\tau\approx 3.3 \times 10^{-15} {\rm s}$ from [4], [5].
(5) holds for the dominant hadronic decay  $A^{-}_{2\pi}\to\pi^{-}\pi^{0}\pi^{0}$ as well as for the negligible electromagnetic decay  $A^{-}_{2\pi}\to\pi^{-}\gamma\gamma$.

	One should note that the vacuum polarisation shift $\Delta E_{VAC}^{-}$ of the ground state energy of $A^{-}_{2\pi}$ cannot be derived from the results on  $P{\! s}^{\!{\! -}}$ or $A_{2\pi}$: it is not proportional to $\rho(\pi^{-}\pi^{+}\pi^{-})$ and it does not depend on only one mass ($m_{c}$), since $m_{e}$ also goes into a calculation of  $\Delta E_{VAC}^{-}$. This shows the limitations of our simple procedure.
On the other hand there is little doubt that 
 \begin{equation}
\Delta E_{VAC}^{-}\approx\Delta E_{VAC},
\label{eq6}
  \end{equation}
where $\Delta E_{VAC}\approx 0.9\; {\rm eV}$ is the vacuum polarisation shift of the ground state of $A_{2\pi}$ [5].
	
	We have anticipated the result  $\frac{2\rho(e^{-}e^{+}e^{-})}{\rho(e^{+}e^{-})}\approx 1$ and will now justify it. For this purpose we use the results of [10], [11], [12]. From [12] we have 
 \begin{equation}
\frac{t}{t^{-}}=\frac{1}{4}\frac{2\rho(e^{-}e^{+}e^{-})}{\rho(e^{+}e^{-})}.
\label{eq7}
  \end{equation}
Here $t$ and $t^{-}$ are the lifetimes of $P{\! s}$ and $P{\! s}^{\!{\! -}}$ respectively; the factor $\frac{1}{4}$ takes into account the fact that the two electrons in  $P{\! s}^{\!{\! -}}$ are in a singlet state, so that the probability of finding one of the electrons in a singlet state with the positron (from which they can annihilate into $\gamma\gamma$) is  $\frac{1}{4}$. From [10] we have $t^{-}=0.478\times 10^{-9} s$, which agrees well with the experiment [11]. Using  $t=1.25\times 10^{-10} {\rm s}$ for the lifetime of $P{\! s}$ from the singlet state, we get
 \begin{equation}
\frac{2\rho(e^{-}e^{+}e^{-})}{\rho(e^{+}e^{-})}\approx 1.05.
\label{eq8}
  \end{equation}
	
	In a second note we will estimate the formation probability of $A^{-}_{2\pi}$ in the experiment [3].

	We thank L. Nemenov, G. C. Oades and W. S. Woolcock for useful discussions and suggestions.

  \end{document}